# Wavelength Spacing Tunable, Multiwavelength Q-switched Fiber Laser Mode-locked by Graphene Oxide

Lei Gao *Student Member, IEEE*, and Tao Zhu *Member, IEEE*

*Abstract*—We demonstrate a wavelength spacing tunable, multiwavelength Q-switched mode-locked fiber laser (QML) based on a fiber taper deposited with graphene oxide. The operation of the laser can be understood in terms of the formation of bunches of QMLs which possess small temporal intervals, and multiwavelength spectra are generated due to the Fourier transformation. We find that the temporal spacing of the QMLs is highly sensitive to the pump power, and as a result, the wavelength spacing can be easily tuned by varying the pump power. Our experimental laser provides a wavelength spacing tuning range from ~0.001 nm to 0.145 nm with a pump power variation less than 10 mW. The laser could be developed into a low lost wavelength spacing tunable optical source for a wide range of applications, such as spectroscopy, microwave/terahertz signal generation, optical metrology, optical communications and sensing.

## I. INTRODUCTION

Wavelength spacing tunable, multiwavelength fiber lasers have potential applications in optical frequency metrology, microwave/Terahertz generation, metrology, sensing, and fiber communications [1-5]. Multiwavelength fiber lasers based on in-fiber comb filters or specific fiber Bragg gratings have been proposed, where wavelength spacing tuning is realized by mechanically controlling the optical path difference or the polarization state of light [6-10]. Nevertheless, the use a rare earth-doped fiber (homogeneous gain medium at room temperature) in such lasers makes it difficult to achieve stable multiwavelength operation because of strong mode competition. Special ways must be employed to overcome this problem, such as cooling gain fiber in liquid nitrogen and adding a polarization-dependent device with spatial hole burning [11, 12]. Besides, the tuning resolution and the repeatability are limited by the precision accuracy of mechanical servo-system. Stable multiwavelength lasers based on nonlinear effects such as four-wave-mixing [13], stimulated Raman scattering [14], stimulated Brillouin scattering and optical parametric oscillation have been demonstrated [15, 16]. However, wavelength spacing tuning in those methods is still inconvenient.

Another practical technique to achieve stable multiwavelength operation is the formation of bound state (BS) of continuous wave mode-locked laser (CWML), namely, the generation of multiple-pulse trains with specific temporal intervals [17-21]. According to the Fourier transformations, a small temporal spacing (in the picosecond/femtosecond range) of the pulses would lead to a giant modulation in the corresponding optical spectrum. As a balance of dispersion and nonlinearity, the BS of CWML has been both theoretically predicted by the complex Ginzburg-Landau equations, and experimentally observed [17-23]. Although it is possible to tune wavelength spacing by varying the BS orders via pump power or polarization controller, the pump power variation required is high (hundreds of mW) and polarization controlling repeatability is poor. There is a need to find a more effective means to tune the wavelength spacing of a multiwavelength fiber laser.

In this paper, we experimentally demonstrate that effective wavelength spacing tuning can be realized in a Q-switched fiber laser (QML), which is formed in an erbium-doped fiber (EDF) cavity that incorporates a fiber taper deposited with graphene oxide (GO). As a two-dimensional layer of hexagonal packed carbon atoms, graphene shows excellent saturable absorption properties, which include broad wavelength range, large modulation depth, low saturation intensity, and its nonlinear susceptibility is $10^8$ times of a commercially available fiber [24-31]. The principle of our multiwavelength laser can be explained by the bunch of QMLs, which is referred as the grouping of several QMLs. Experiments show that this laser can offer a wavelength spacing tuning from ~0.001 nm to 0.145 nm with a pump-power variation less than 10 mW.

## II. PRINCIPLE of OPERATION

In contrast with CWML, a QML shows a giant Q-switched envelope above the pulse trains with two characteristic repetition frequencies in its frequency spectrum, and the period and the width of the pulse envelope are sensitive to the pump power [32-35]. Based on various nonlinear elements, such as nonlinear polarization rotators [34], single-wall carbon nano-tubes [35], semiconductor saturable absorber mirrors [36], and graphene [26, 32], a variety of schemes have been proposed for QML. Figure 1 (a) denotes schematically the typical characteristics of a QML, where two parameters, $t_1$ the period of the envelope, and $t_2$ the period of the pulses, are defined, and $f_1$ and $f_2$ are the corresponding quantities in the frequency domain, respectively. In the whole pumping region, only $t_1$ varies with the pump intensity, and $t_2$ keeps constant. In the optical spectrum, both the $f_1$ and $f_2$ cannot be detected due to the limited resolution of a conventionally optical spectrum analyzer (OSA).

When several QMLs are bunched together in a laser system, the total optical field of the laser train, $P(t)$, can be described as

$$P(t)=\sum_{m=1}^{m=N} P_{mQML}(t) \quad (1)$$

where $N$ is the number of the QMLs that bunched together, and $P_{mQML}(t)$ represents the optical field of $m$th QML. In

This work was supported by Natural Science Foundation of China under Grant 61377066, and the Fundamental Research Funds for the Central Universities under Grants CDJZR12125502 and 106112013CDJZR120002, 106112013CDJZR160006. Discussions with Prof. Zhipei Sun from Aalto University are appreciated.
All authors are with the Key Laboratory of Optoelectronic Technology & Systems (Ministry of Education), Chongqing University, Chongqing, 400044, China. (*Corresponding email: zhutao@cqu.edu.cn* ) some cases, although more than two QMLs are bunched together, only the temporal spacing needs to be considered. This assumption is valid when the wavelength spacing is larger than 0.1 nm, corresponding to a temporal interval less than ~100 ps. In this case, the profiles of the QMLs can be rationally assumed as the same, and Eq. (1) can be simplified

$$P(t)=[P_{QML}(t)+P_{QML}(t)e^{j\omega\Delta t}e^{j\Delta\phi}] \qquad (2)$$

where $\omega$ is the laser angular frequency, $\Delta t$ is the temporal spacing of the two QMLs, and $\Delta \Phi$ is their phase difference. The corresponding optical spectrum $P(\omega)$ can be calculated by taking Fourier transform of $P(t)$,

$$P(\omega)=2P_{QML}(\omega)e^{-j(\frac{\omega}{2}\Delta t-\frac{\Delta\phi}{2})} \cdot \cos(\frac{\omega}{2}\Delta t-\frac{\Delta\phi}{2}) \qquad (3)$$

Thus, the frequency intensity $I(\omega)$ should be

$$I(\omega)=2|P(\omega)|^2 \cdot [1+\cos(\omega\Delta t-\Delta\phi)] \qquad (4)$$

That's to say, when two QMLs form a bunch, the original optical spectrum is modulated by a frequency $1/\Delta t$, which is inversely proportional to the temporal interval. In our experiment, the $\Delta t$ in Eq. (4) varies under different pump powers, and we define four regions depending on wavelength spacing characteristic shown in its corresponding optical spectrum, wavelength spacing increasing (WSI) region, single wavelength (SW) region, wavelength spacing decreasing (WSD) region, and ultra-dense wavelength spacing (UWS) region. The pump laser is increased continuously in the four regions, and $t_3$, the temporal period of the bunch of QMLs, keeps decreasing.

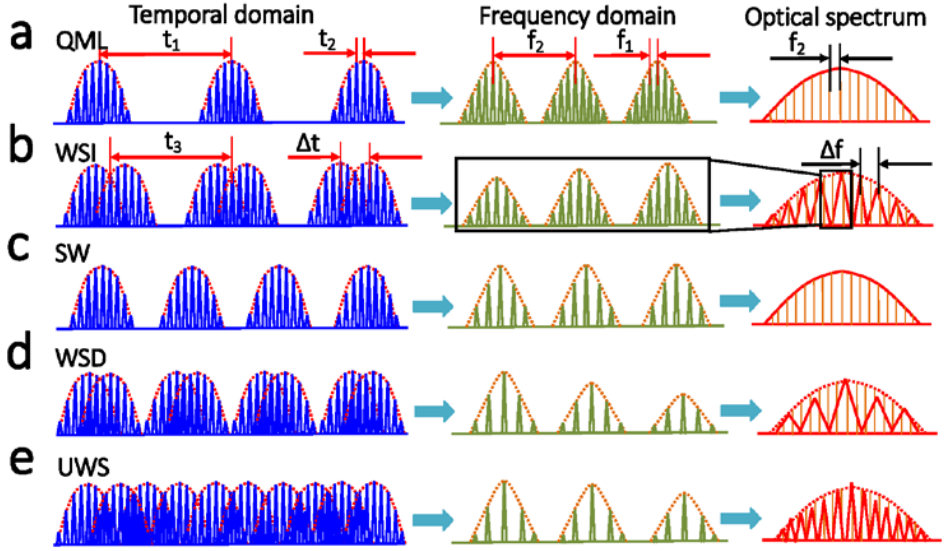

Fig.1. Schematic characteristics of (a) a typical QML; (b) WSI region; (c) SW region; (d) WSD region, and (e) UWS region.

The WSI region is schematically shown in Fig. 1 (b), where $\Delta t$ keeps decreasing until it reaches zero. Consequently, a modulation is shown in its optical spectrum, and the period of this modulation keeps increasing in this region. The SW region emerges when $\Delta t$ is zero, and $\Delta f$ should be infinite in theory. Yet, the SW region is hard to be observed experimentally as it is very sensitive to the pump intensity. The WSD region that requires higher pump powers are represented in Fig. 1. (d), where $t_3$ keeps decreasing with $\Delta t$ increased in the whole region. In the UWS region, where $t_3$ is small enough, the optical spectrum is too dense to find out the details, but this region can be identified by its corresponding temporal trains.

III. EXPERIMENTAL SETUP

The scheme of fiber ring cavity is shown in Fig. 2 (a). 10 m long EDF (EDF-980-T2) is pumped by a 980 nm laser through a wavelength division multiplexer (WDM). A polarization-independent isolator is used for unidirectional operation, and a polarization controller (PC) is employed to change the birefringence of the cavity. The output laser is extracted from a 10% fiber coupler. The saturable absorber (SA) is produced by depositing graphene oxide on fiber taper via light pressure. Compared with previous reports [28], the 5 mm taper waist in our experiment provides long evanescent field interacting length between the GO and the laser, guaranteeing excellent saturable absorption and high nonlinearity. Besides, the large loss of SA facilitates the laser function in Q-switched mode-locking. As no polarization-dependent component is used, mode-locking based on nonlinear polarization rotation would not happen in this cavity when the SA was absent. Other fibers used in the cavity are standard SMFs (SMF-28). The total length of the ring cavity is about 24.5 m, corresponding to a fundamental frequency of ~8.4 MHz. The group velocity dispersions (GVD) parameters of EDF and SMF are -12.7 and 18 ps/nm/km, respectively, implying that the net dispersion is anomalous.

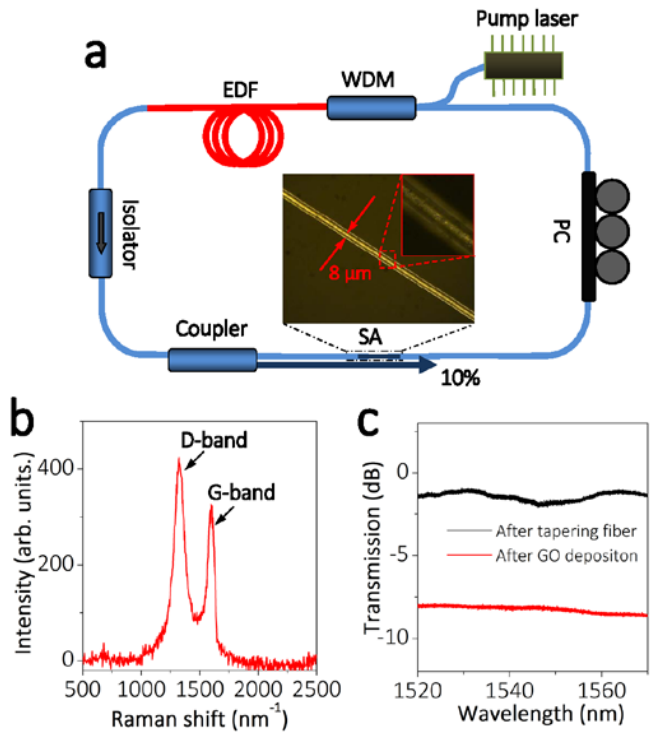

Fig.2. (a) Schematic of the fiber ring cavity; (b) Raman spectrum of the GO; (c) Transmission spectra of the fiber taper before and after GO deposition.

The GO used in our experiment is prepared by chemical oxidation of graphite. 0.5 mg GO is immersed into 100 mL N,N-dimethylformamide solution, and after 20 minutes of ultra-sonication, the solution is centrifuged to get uniform and transparent layer. The Raman spectroscopy is shown in Fig. 2 (b), where D-band and G-band appear at around 1319 cm$^{-1}$ and 1606 cm$^{-1}$, respectively. The ratio of $I_D/I_G$ of 1.3 indicates that the GO has a large amount of defects [37]. The fiber taper is produced from the standard SMF with good repeatability. The waist diameter of the taper is ~ 8 μm, and the waist length is ~5 mm. The results shown in Fig. 2 (c) indicate an taper insertion loss of 1.5 dB. After immersing the fiber taper into a droplet of transparent GO solution, a 200 mW continuous wave 980 nm laser is injected into one port of the taper, and the output power is monitored by an optical power meter. Two minutes later, the GO deposition loss is ~7 dB, and the taper is removed from GO solution and fixed in a clean box for natural drying. Such a high deposition loss guarantees enough GO flakes to function with the evanescent field. The transmission spectrum in Fig. 2 (c) shows no absorption peak. The microscopic image of the fiber taper after GO deposition shown in Fig. 2 (a) indicates that GO flakes have been well deposited on the taper waist.

The temporal output of laser is monitored by a detector (PDB430C, 350MHz, Thorlabs Co,. Ltd) and visualized by a real time oscilloscope (Infiniium MSO 9404A, Agilent Tech.) and a frequency analyzer (DSA815, Rigol. Corp.). An OSA (B6142B, Advantest Corp.) is utilized to measure its optical spectra. To ensure the detector is not saturated, a variable optical attenuator is inserted between the detector and output fiber. An OSA (Si720, Micro Optics) is utilized to monitor the transmission spectrum of the fiber taper during the fiber-tapering and GO-deposition process. The Raman spectrum of the GO is excited with a 632.8 nm laser by an home-made Raman spectrometer.

## IV. EXPERIMENTAL RESULTS

Increasing pump power from 0 to 380 mW with a rate of ~1 mW, multiwavelength QML is observed. As the pulse envelope is so big and more QMLs are grouped in one total pulse, the resolution and range requirements cannot be satisfied by an autocorrelator at the same time. Here we only give the results recorded by a photodetector with a bandwidth of 350 MHz. The typical temporal train of multiwavelength QML is shown in Fig. 3 (a), which posses a giant envelope above short pulse trains, and part of it shown in Fig. 3 (b) indicates that the laser pulse contains more than 3 QMLs, and each of the QML is formed by a number of pulse groups containing 9 ML pulses spacing with the cavity length. That means each single QML is modulated constantly in the whole experiment, and several QMLs formed a bunch of QMLs. Its RF spectrum in Fig. 3 (c) contains two repetition frequencies of 25.6 kHz and 8.42 MHz that corresponds to the total pulse period and the period of ML pulse, respectively, and a peak-to-background ratio of ~60 dB in the inset indicates that the laser has been mode-locked well. Limited by the bandwidth of the frequency analyzer, the modulation period of the 9 ML pulses is not shown. Its optical spectrum in Fig. 3 (d) indicates a wavelength spacing of ~0.015 nm, corresponding to a bandwidth of 1.87 GHz, which is larger than that of our detection system, so the actual temporal spacing of the QMLs is not shown.

Figure 4 depicts the spectra with different pump powers, where the wavelength spacing increases when pump power is below ~53 mW, but decreases when further increasing pump power. The four processes of bunches of QML under different pump intensities are shown in Fig. 5 (a), where WSI appears once the laser is formed, then WSD and UWS are observed with higher pump powers. As the formation of bunches of QMLs is very sensitive to the pump intensity, SW is not observed due to the limited step of the pump laser. The center wavelength red-shifts for larger pump power, and its full width at half maximum (FWHM) increases linearly with the increment of pump power. The pulse width and pulse repetition rate are plotted as a function of pump power in Fig. 5 (b), where the repetition rate increases linearly from 16.8 kHz to 37.3 kHz and pulse width decreases from 41.5 μs to 7 μs when pump power increases from 20 mW to 76 mW. The wavelength spacing of the laser vs pump strength is plotted in Fig. 5 (b). In our experiment, a maximum wavelength spacing of 0.145 nm is obtained, corresponding to a minimum wavelength number of 7. However, larger wavelength spacing, or even the SW region, is rationally expected to be observed when pumped with a smaller power step. For pump power larger than 76.2 mW, the laser cavity functions in UWS region.

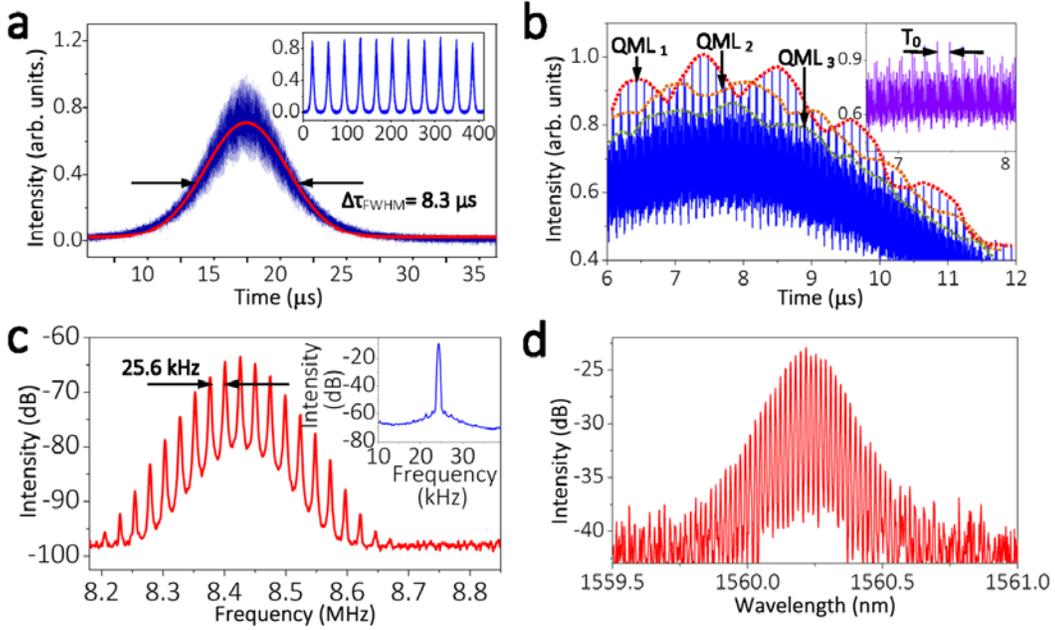

Fig.3. Characteristics of a typical bunch of QMLs. (a) Temporal pulse train with pump power at 47 mW, fitted with Gauss profile, the inset shows the pulse train in a large range; (b) Part of a laser pulse, the inset shows the ML pulses spacing with the cavity length; (c) Corresponding RF spectrum and (d) Optical spectrum.

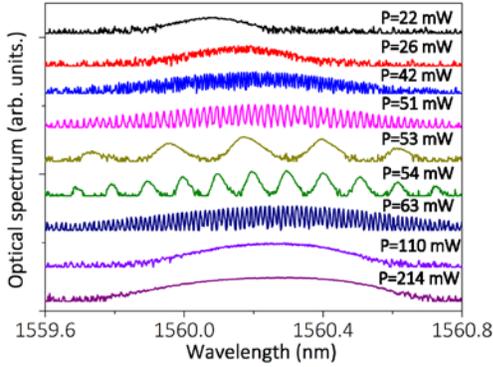

Fig.4. Optical spectra at different pump powers.

Limited by the maximum bandwidth of our system, the temporal trains are obscure for large wavelength spacing, however, when the spacing is relatively small, the difference of their temporal outputs can be figured out. As shown in Figs. 6 (a) and (b), the spacing of two QMLs is about 13 ns, so the fine structure cannot reveal in the optical spectrum. When the wavelength spacing is ~ 0.002 nm in Figs. 6 (c), the temporal spacing of ~1.9 ns is shown in its corresponding trains. As in UWS, no fine structure can be found in the optical spectrum, as represented in Figs. 6 (e) and (f).

## V. DISCUSSIONS

Different from previous reports [26, 38], no continuous wave laser ever was shown, and QML emerged other than CWML as long as the laser is formed. The former is mainly induced by the low saturation energy of the SA, while the later is that a minimum intra-cavity energy to completely bleach the SA is required to get CWML fiber laser [33], and the larger loss of SA also expedites the tendency toward QML. Previously, self-mode-locking has been observed in passively Q-switched lasers due to intermodal beats, where both a single pulse and pulse trains are modulated by a fundamental cavity frequency [39, 40]. Yet, results in our experiment are different from observations in both the temporal output and optical spectrum. Besides, the fine structures in the laser trains illustrate that they are formed because of the bounding of pulses, not interference effect.

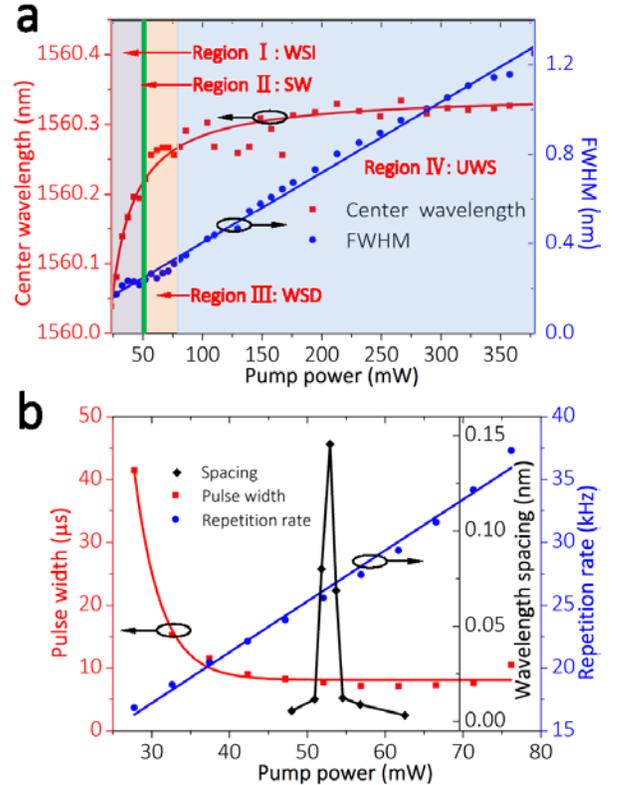

Fig.5. Pump-dependent characteristics of bunches of QMLs. (a) The center wavelength and the FWHM of lasing wavelength vs pump power. The center wavelength is fitted exponentially while FWHM is fitted linearly, and the envelope is counted for the multiwavelength region; (b) Temporal pulse width, repetition rate and wavelength spacing vs pump power. The pulse width is fitted exponentially while repetition rate is fitted linearly.

Compared with CWML, QML posses a more sophisticated mechanism as more physical parameters are involved [33], and a clear physical interpretation of the formation of bunches of QMLs is unknown. During the experiment, we find that the formation of bunches of QMLs requires a delicate balance of dispersion, nonlinearity, insertion loss and most of importance, pump strength. In our experiment, it can be only found in cavity with net anomalous GVD, of which is verified by splicing different lengths of SMF and dispersion compensating fiber into the cavity to change its net dispersion.

A proper insertion loss of the SA, including both taper loss and deposition loss of GO, is very important in generating BS of QML. Considering that QML is formed under low pump strength, it is not surprising to find that bunches of QMLs are extremely sensitive to pump power. This intensely pump-dependent characteristic can be easily understood that energy in the laser cavity is insufficient for completely bleaching the SA, so a slight variation of pump power would lead to a great change of mutual forces between the laser pulses.

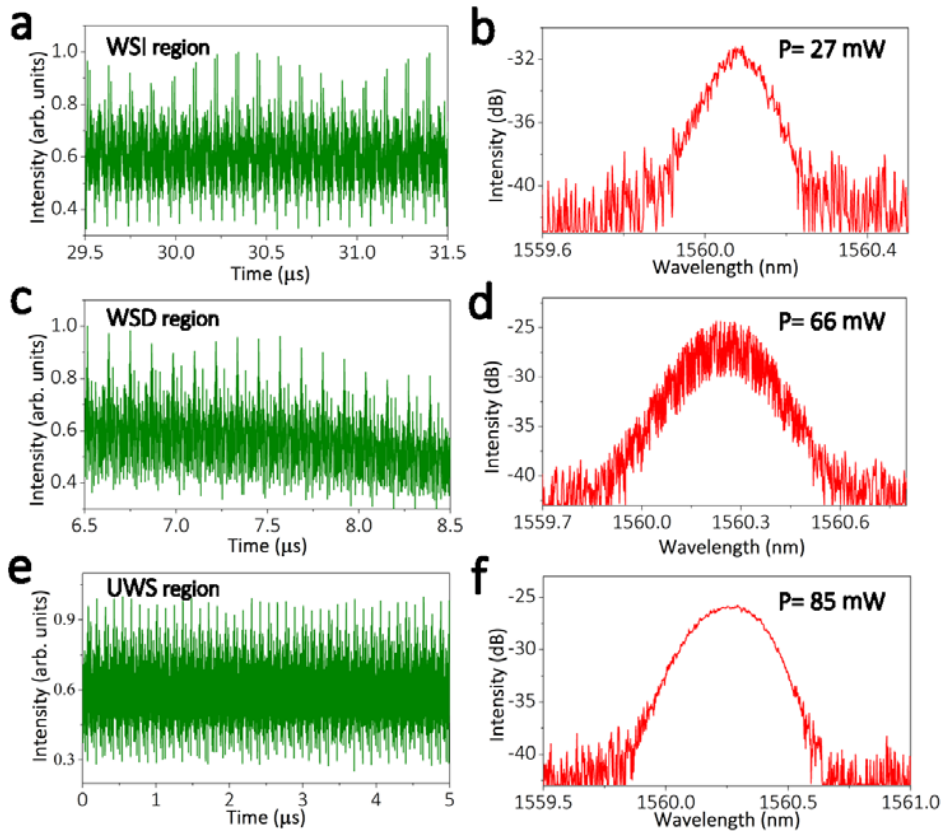

Fig.6. Temporal trains and the corresponding spectra of bunches of QMLs in the three regions. (a) and (b) are for WSI region, (c) and (d) are for WSD region, (e) and (f) are for UWS region.

## VI. CONCLUSIONS

We have proposed a wavelength spacing tunable, multiwavelength Q-switched mode-locked fiber laser in an EDF cavity based on a fiber taper deposited with graphene oxide. Due to the Fourier transformation between the temporal and frequency domains, the formation of bunch of QMLs leads to multiwavelength spectrum. As a result of delicate balance of dispersion, nonlinearity, insertion loss and pump strength, the temporal spacing of the QMLs is highly sensitive to the pump power, so the wavelength spacing can be easily tuned by varying the pump power. Experimental results show that the wavelength spacing can be tuned from ~0.001 nm to 0.145 nm with a pump-power variation less than 10 mW, and it can be even larger if a pump laser with a smaller intensity step was utilized. To our best knowledge, this is the first report about experimental observation of bunches of QMLs, and it also provides a new mechanism to fabricate tunable multiwavelength laser. This kind of laser is beneficial to understanding the complex dynamics of laser physics, and also would find potential applications in optical communication, sensing, spectroscopy, optical metrology, and microwave/terahertz generation.

**Lei Gao** was born in Henan, China, in 1989. He received the B.S. degree in Optoelectronic information engineering from Chognqing University, Chongqing, China, in 2011. He is currently working toward the PhD. degree in optical engineering from the Key Laboratory of Optoelectronic Technology and Systems, Ministry of Education, Chongqing University, Chongqing, 400044, China. His current research interests include fiber laser and photonics.

Mr. Gao is a student member of the IEEE and the Optical Society of America.

**Tao Zhu** (M'06) received the Ph. D. degrees in Optical Engineering from Chongqing University, Chongqing, China, in 2008. During 2008-2009, he worked in Chongqing University, China. During 2010-2011, he is a Postdoctoral Research Fellow at the Department of Physics in University of Ottawa, Canada. Since April 2011, he is a professor of Chongqing University, China. He has published over 80 papers in the international journals and the conference proceedings. His research focuses on passive and active optical components, optical sensors, and distributed optical fiber sensing system.

Dr. Zhu is a member of the IEEE and the Optical Society of America.